\documentclass[letterpaper, 10 pt, conference]{ieeeconf}  

\IEEEoverridecommandlockouts                              
\usepackage[utf8]{inputenc}
\usepackage[T1]{fontenc}

\title{\LARGE \bf
Prospect Theoretic Approach to Pursuit-Evasion Differential Games with Risk Aversion and Probability Sensitivity
}

\author{Zili Wang$^{1}$, Hao Yang$^{1}$ and Xiangxiang Wang$^{1}$
\thanks{*This work was supported by the National Natural Science Foundation of China (62233009, 62625305) and Basic Research Program of Jiangsu (BK20253024). Corresponding author: Hao Yang.}
\thanks{$^{1}$Zili Wang, Hao Yang and Xiangxiang Wang are with the College of Automation Engineering, Nanjing University of Aeronautics and Astronautics, Nanjing 210016, China (e-mail: wangzili@nuaa.edu.cn; haoyang@nuaa.edu.cn; xiangxiangwang@nuaa.edu.cn).}}

\usepackage{cite}
\usepackage{amssymb,amsfonts}
\usepackage{algorithmic}
\usepackage{graphicx}
\usepackage{algorithm,algorithmic}
\usepackage{hyperref}
\hypersetup{hypertex=true,
            colorlinks=true,
            linkcolor=blue,
            anchorcolor=blue,
            citecolor=blue}
\hypersetup{hidelinks=true}
\usepackage{textcomp}
\usepackage{xcolor}
\usepackage{stfloats}
\usepackage{url}
\usepackage{verbatim}
\usepackage{graphicx}
\usepackage{amssymb}
\usepackage{amsmath,bm}
\usepackage{mathrsfs}
\usepackage{arydshln}
\usepackage{nicematrix}
\let\labelindent\relax
\usepackage{enumitem}
\usepackage{tikz}
\usetikzlibrary{arrows.meta,positioning}
\newlength{\caseLabelWidth}
\newtheorem{theorem}{Theorem}
\newtheorem{lemma}{Lemma}
\newtheorem{proposition}{Proposition}
\newtheorem{corollary}{Corollary}

\newtheorem{definition}{Definition}
\newtheorem{assumption}{Assumption}
\newtheorem{remark}{Remark}
\newtheorem{problem}{Problem}
\providecommand{\qedsymbol}{\square}

\newenvironment{IEEEproof}{%
  \par\smallskip\noindent\textit{Proof:}\enspace
}{%
  \par\smallskip
}

\newlist{caselist}{enumerate}{1}
\setlist[caselist,1]{%
  label=\textbf{Scenario~\arabic*.},
  leftmargin=*,                 
  labelwidth=\caseLabelWidth,   
  labelsep=0.2em,              
  align=left,
  itemsep=0pt,
  topsep=0pt,
  parsep=0pt,
  partopsep=0pt
}
\def\BibTeX{{\rm B\kern-.05em{\sc i\kern-.025em b}\kern-.08em
    T\kern-.1667em\lower.7ex\hbox{E}\kern-.125emX}}
    
\begin{document}

\maketitle
\thispagestyle{empty}
\pagestyle{empty}

\begin{abstract}                          
This paper considers pursuit-evasion (PE) differential games in which the pursuer's and evader's irrational perceptions of environmental uncertainty lead to deviations from rational controllers. Firstly, the equivalent irrational behavioral models of risk aversion and probability sensitivity are established using Cumulative Prospect Theory (CPT). Secondly, sufficient conditions for capturability are established in terms of irrationality parameters. Finally, the CPT-Nash equilibrium is analyzed by Brouwer’s fixed-point and self-consistency theorem. The new results reveal that irrational behaviors benefit the pursuer in some cases and the evader in others. Certain captures that are unachievable under rational behaviors can be achieved under irrational ones. By bridging irrational behavior with game-based control, a theoretical foundation is established for human-machine control systems.
\end{abstract}

\section{Introduction}
Pursuit-evasion (PE) differential game is an important branch of differential games where the pursuer and the evader with opposing objectives control their motion over time to minimize or maximize a certain performance index \cite{isaacs1965differential,lopez2020solutions}. The capturability and Nash equilibrium problems have also been investigated using various methods \cite{Liu2025Distributed,yan2022matching,ye2023distributed,Borkar2021SCL}.

In practical situations, environmental changes often alter the relative priorities of different objectives (efficiency, safety, etc.) that need to be optimized \cite{zhang2021pursuer,Li2026SCL}. Such uncertainty is often described by a Bayesian differential game where the uncertain weights are represented by probability distributions that are known a priori \cite{lopez2020bayesian}.

Most studies on Bayesian differential games assume "rational" players who make optimal decisions based on objective probability. Nevertheless, subjective bias in perceiving probabilities has not been taken into account. Such irrationality widely exists in human-machine interactions \cite{hobbs2018early,wu2022survey,Kanellopoulos2019SCL}, where humans perceive the probability distributions with \textit{risk aversion} and \textit{probability sensitivity} \cite{kahneman1979prospect}. Thus, humans tend to avoid perceived performance loss, and underestimate large probabilities (or, correspondingly, overestimate small probabilities). Empirical evidence from social experiments indicates that these two irrational behaviors are important sources of deviations between human decisions and optimal outcomes under rationality~\cite{camerer2011behavioral,tversky1992advances}.

Hence, it is desirable to incorporate irrational perception into Bayesian PE differential games. However, directly applying existing frameworks to the irrational case is challenging, since irrational players no longer optimize the original performance indices. Therefore, an equivalent representation of subjective performance evaluations is needed so that their effects on feedback controllers can be analyzed.

On the other hand, Prospect Theory \cite{kahneman1979prospect}, which led to a Nobel prize achievement for Kahneman, provides a comprehensive framework involving risk aversion and probability sensitivity. This theory has been widely applied to economics and social sciences \cite{li2024imperfect}. Currently, in engineering applications, Cumulative Prospect Theory (CPT) has been incorporated into games to enable the mathematical characterization of perceived gain/loss \cite{Hota2019Ga,sanjab2020game}. Although CPT-based optimization has been numerically solved \cite{jie2018stochastic}, extending CPT from single optimization to PE differential games is still a key problem. A main challenge lies in the limited understanding of how to model the irrational perception, and how to analyze the capturability and Nash equilibrium.

This work focuses on a fundamental question: \emph{how do irrational perceptions of risk aversion and probability sensitivity affect controllers, capturability and Nash equilibrium?} The main contributions are summarized as follows:

1. To describe subjective bias in perceiving the environmental uncertainty in practical human-machine control systems, an equivalent behavioral performance model is incorporated via a general CPT function. The irrational behaviors, including loss aversion, overweighting small probabilities, and underweighting large probabilities are characterized.

2. By bridging irrational behavior with game-based control, a theoretical foundation is established, where capturability is analyzed for three scenarios with different PE capabilities. Sufficient capturability conditions are rigorously established in terms of CPT irrationality parameters (see Theorem~1). The existence of the CPT-Nash equilibrium is also analyzed, which can be guaranteed by invoking Brouwer's fixed point theorem (see Proposition~3) and self-consistency analysis (see Proposition~4).

3. The results show that the optimal behavior not only depends on the distance and control consumption but also on the CPT function (see Proposition~2). Irrational behaviors benefit the pursuer in some cases and the evader in others. Certain captures that are unachievable under rational behaviors can be achieved under irrational ones. This provides a solid foundation for regulating the effect of irrational behavior in PE differential games.

The remainder of the paper is organized as follows: Section 2 establishes the model for irrational perceptions. Section 3 gives the main result. The proof of the main result is shown in Section 4. Numerical examples are presented in Section 5, followed by a conclusion in Section 6.

\textit{Notations:} We denote by $x^{T}$ (or $A^{T}$) the transpose of a vector $x$ (or a matrix $A$). The notation $X\geq 0$ ($X>0$) indicates that $X$ is a positive semi-definite (definite) matrix. Moreover, $I_{n\times n}\in\mathbb{R}^{n\times n}$ denotes the identity matrix.

\section{Model of Irrational Perceptions}
In this section, the Bayesian PE differential game is introduced in 2.1 and the model of the CPT-based PE differential game is established in 2.2.
\subsection{Bayesian PE Differential Game}
Consider the PE differential game with a single pursuer and a single evader, whose dynamics satisfy
\begin{equation}
\label{agent dynamic}
\left\{\begin{array}{l}
\dot{z}_1= B_1 u_r, \;\; z_1(0)=z_{1 0}\\
\dot{z}_2= B_2 v_r, \;\; z_2(0)=z_{2 0}
\end{array}\right.
\end{equation}
where $B_i$ is of full rank, $z_i \in \mathbb{R}^{3}$ are the states for $i \in \{1,2 \}$, and $u_r \in \mathbb{R}^3$, $v_r \in \mathbb{R}^3$ are the control inputs of the pursuer and evader, respectively. The system (\ref{agent dynamic}) can be rewritten as
\begin{equation}
\label{dynamic}
\dot{x}(t)= B_1 u_r- B_2 v_r, \;\; x(0)=x_0
\end{equation}
where $x \triangleq z_1-z_2$, $x_0 \triangleq z_{10}-z_{20} \neq 0$.

The goal of the pursuer is to minimize the PE distance and energy consumption, while the evader desires to maximize the distance and minimize energy consumption. The performance index is defined as
\begin{equation}
\label{nominal performance index}
J_r(x_0, u_r, v_r) \triangleq \int_0^{\infty} ( x^T Q_r x + u_r^T R u_r -v_r^T \Pi v_r) d \tau
\end{equation}
where $Q_r \in \mathbb{R}^{3 \times 3}$ is the positive definite state reward weighting matrix, $R \in \mathbb{R}^{3 \times 3}$ and $\Pi \in \mathbb{R}^{3 \times 3}$ are the positive definite control input reward weighting matrices.

According to optimal control theory \cite{anderson2007optimal}, the optimal controller can be expressed as
\begin{equation}
\label{nominal controller}
\begin{aligned}
u_r^*=- R^{-1} B_1^T P x, \;\;v_r^*=- \Pi^{-1} B_2^T P x 
\end{aligned}
\end{equation}
where the matrix $P$ solves the algebraic Riccati equation
\begin{equation}
\label{nominal riccati equation}
Q_r-P(S_1-S_2)P=0
\end{equation}
with $S_1 \triangleq B_1 R^{-1} B_1^T$, $S_2 \triangleq B_2 \Pi ^{-1} B_2^T$. Capture is defined as follows.

\begin{definition}[\bf{Capture Achievement}] In the PE differential game (\ref{dynamic})-(\ref{nominal performance index}), the capture can be achieved if $\lim\limits_{t \rightarrow \infty} x(t)=0$ for any admissible optimal controller (\ref{nominal controller}).~$\hfill \square$\end{definition}

\begin{lemma}[\bf{\cite{basar1995dynamic}}] Consider the PE differential game (\ref{dynamic})–(\ref{nominal performance index}) with the optimal controllers (\ref{nominal controller}) satisfying (\ref{nominal riccati equation}). The capture is achieved if and only if 
\begin{equation}
\label{lemma}
B_1 R^{-1} B_1^T -B_2 \Pi^{-1} B_2^T >0
\end{equation}
is satisfied. $\hfill \blacksquare$ \end{lemma}

The uncertainty arising from objective environmental fluctuations leads to the probabilistic nature of the weight matrix $Q(\omega)$ \cite{sun2023stochastic}. The resulting performance index becomes
\begin{equation}
\label{probabilistic performance index}
J(x_0, u, v, \omega) \triangleq \int_0^{\infty} ( x^T Q(\omega) x + u^T R u -v^T \Pi v) d \tau
\end{equation}
where
\begin{equation}
\label{Q probability}
Q(\omega) = Q_r + \eta(\omega) \Sigma_q
\end{equation}
with the uncertainty-scale matrix $\Sigma_q \triangleq q Q_{\Delta}$, $Q_{\Delta} > 0$, constant $q>0$ and $\eta \sim \mathcal{N}(0, 1)$. For any admissible controllers independent of $\eta$, it follows that
$$
J(x_0,u,v,\omega) =J_r(x_0,u,v)+\eta(\omega)\int_0^\infty x^T\Sigma_qx\,d\tau
$$

The expectation of $J$ can be written as
\begin{equation}
\label{mathematical expectation}
\begin{aligned}
\mathbb{E}[J] & \!= \!\mathbb{E}\big[\!\int_0^{\infty}\!x^T Q(\omega) x d\tau \big] \!+\!\!\int_0^{\infty}\!(u^TRu-v^T \Pi v) d\tau \\
& = \int_0^{\infty}(x^T Q_r x+u^TRu-v^T \Pi v) d\tau.
\end{aligned}
\end{equation}

Denote $\sigma \triangleq \int_0^{\infty} x^T \Sigma_q x d\tau$, it follows that
\begin{equation}
\label{Var J}
\operatorname{Var}[J] = \operatorname{Var}[J_r] +\operatorname{Var} \big(\int_0^{\infty} \eta(\omega) x^T \Sigma_q x d \tau \big) = \sigma^2.
\end{equation}
It is an obvious observation that
$$J(x_0,u,v,\omega) \sim \mathcal{N}(J_r(x_0,u,v), \sigma^2).$$

\begin{assumption} The pursuer and evader know the common prior distribution of the weighting matrix $Q(\omega)$. $\hfill \qedsymbol$ \end{assumption}

Then the capturability condition of Lemma 1 can be extended to the Bayesian setting in the following Lemma:
\begin{lemma}Under Assumption 1, consider the Bayesian PE differential game (\ref{dynamic}), (\ref{probabilistic performance index}), with the optimal controllers (\ref{nominal controller}) satisfying (\ref{nominal riccati equation}). Then the capture is achieved if and only if condition (\ref{lemma}) in Lemma 1 is satisfied. \end{lemma}

\begin{IEEEproof} The pursuer and evader determine controllers by minimizing and maximizing, respectively, the performance index $\mathbb{E}[J]$. It can be observed from (\ref{mathematical expectation}) that the expected performance index equals the original reference performance index $J_r$. Consequently, for the random $Q(\omega)$ in (\ref{Q probability}), the Nash equilibrium admits the same controller (\ref{nominal controller}). $\hfill \blacksquare$
\end{IEEEproof}

\subsection{CPT-Based PE Differential Games}
\begin{figure}[htbp]
    \centering
    \centering
    \includegraphics[width=\linewidth]{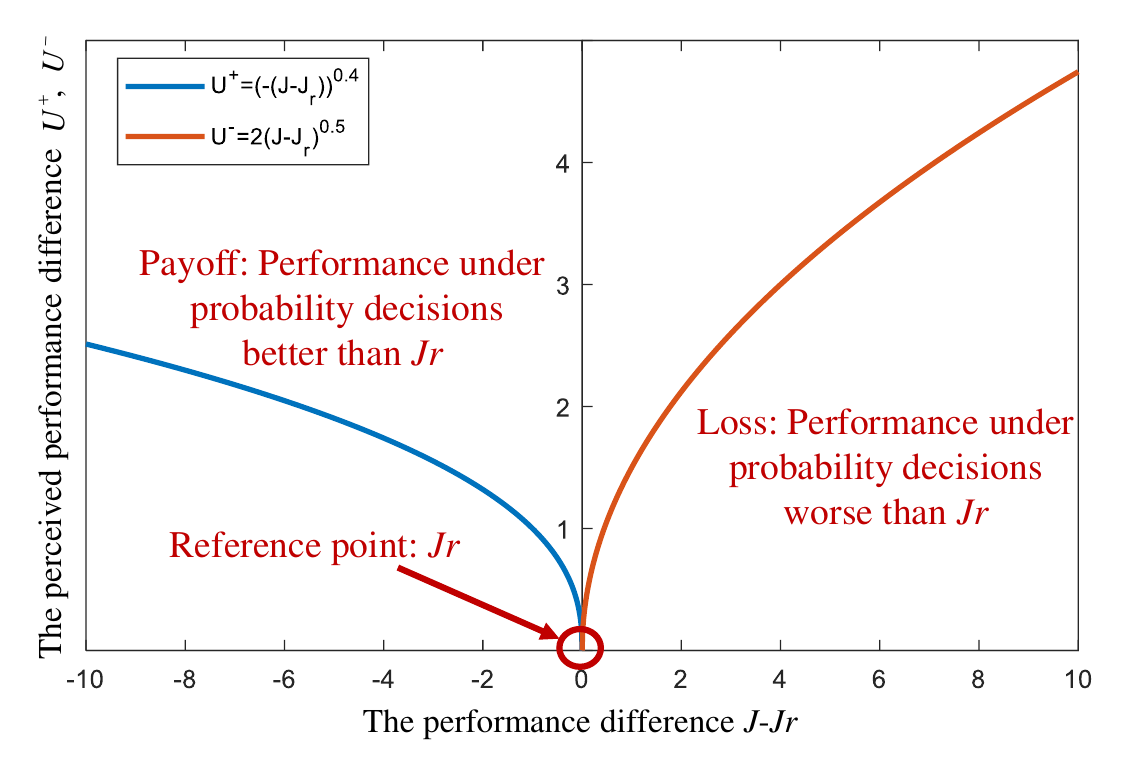}
    \caption{Risk aversion of pursuer}
    \label{Risk aversion}
\end{figure}

This subsection introduces CPT into the Bayesian PE differential game. The corresponding prospect value of $J(u, v)$ can be evaluated by a “CPT-function” that characterizes the risk aversion and probability sensitivity \cite{jie2018stochastic}. The definition is given as
\begin{equation}
\label{CPT function}
\begin{aligned}
\mathbb{C}_i(J)\triangleq & \underbrace{\int_0^{\infty} w_i^{+}\Big(p_i\big(U_i^{+}(J)>h\big)\Big) d h}_{\mathbb{C}_i^{+}} \\
& -\underbrace{\int_0^{\infty} w_i^{-}\Big(p_i\big(U_i^{-}(J)>h\big)\Big) d h}_{\mathbb{C}_i^{-}}
\end{aligned}
\end{equation}
where $\mathbb{C}_i^{+}$ and $\mathbb{C}_i^{-}$, $i = 1,2$ represent the expected positive and negative prospects respectively. The prospect function is explained as:

\textbf{Pursuer} ($i$ = 1): a. The positive prospect $\mathbb{C}_1^{+}$ contains two entities: the gain $U_1^+(J)$ and the perceived probability $w_1^+ (p_1)$. As shown by the blue line in Fig.~\ref{Risk aversion}, $U_1^+(J)$ measures the subjective gain value when the outcome $J$ is better than the reference point $J_r$ (i.e., $J < J_r$, since the pursuer aims to minimize the performance index), defined by
\begin{equation}
\label{U1+}
U_1^{+}(J) \triangleq \big(-(J-J_r)\big)^{\alpha_1}, \;\; J \leq J_r
\end{equation}
where $\alpha_1 \in (0,1]$ represents the gain sensitivity.

\begin{figure}[htbp]
        \centering
        \includegraphics[width=\linewidth]{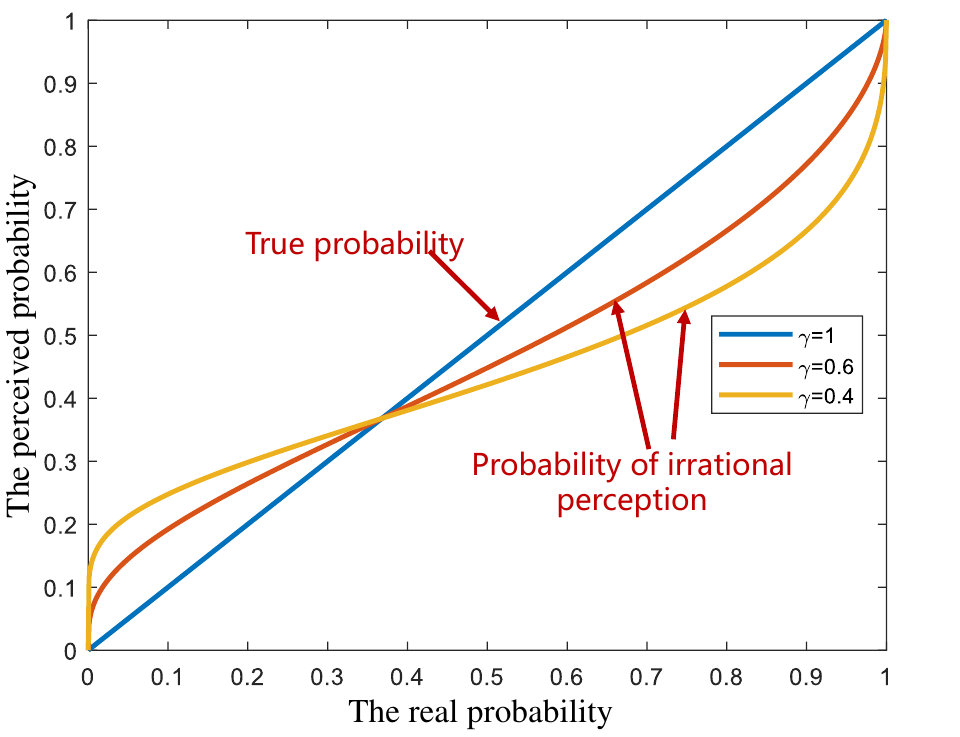}
        \caption{Probability sensitivity of pursuer}
        \label{Probability sensitivity}
\end{figure}

As shown in Fig.~\ref{Probability sensitivity}, $w_1^+ (p_1)$ quantifies the degree of overweighting low probabilities and underweighting high probabilities. The definition is followed by
\begin{equation}
\label{w1}
w_1^{+}(p_1) \triangleq \exp \big\{-\big(-\log (p_1)\big)^{\gamma_1}\big\}
\end{equation}
where $p_1$ is the probability assigned by the pursuer, $\gamma_1 \in (0, 1]$ characterizes the pursuer's perceptual distortion.

b. The negative prospect $\mathbb{C}_1^{-}$ contains the loss $U_1^-(J)$ and the corresponding perceived probability $w_1^- (p_1)$. As shown by the red line in Fig.~\ref{Risk aversion}, $U_1^-(J)$ measures the subjective loss value when the game outcome $J$ is worse than the reference point $J_r$ (i.e., $J > J_r$), defined by
\begin{equation}
\label{U1-}
U_1^{-}(J) \triangleq \epsilon_1(J-J_r)^{\beta_1}, \;\; J>J_r,
\end{equation}
where $\beta_1 \in (0,1]$ represents the loss sensitivity and $\epsilon_1 \in [1,\infty)$ is known as the loss multiplier. Therefore, $\mathbb{C}_1^{-}$ represents the expected loss that makes the performance of pursuer worse.

\textbf{Evader} ($i$ = 2): a. The positive prospect $\mathbb{C}_2^{+}$. $U_2^+(J)$ represents the gain that makes $J$ larger than $J_r$, defined by
\begin{equation}
\label{U2+}
U_2^{+}(J) \triangleq (J-J_r)^{\alpha_2}, \;\; J\geq J_r,
\end{equation}
where $\alpha_2 \in (0,1]$ shares the same meaning as $\alpha_1$. The probability weighting function is then defined by
\begin{equation}
\label{probability weighting function2}
w_2^{+}(p_2)\triangleq \exp \big\{-\big(-\log (p_2)\big)^{\gamma_2}\big\}
\end{equation}
with the probability $p_2$ assigned by the evader, $\gamma_2 \in (0, 1]$. $\mathbb{C}_2^+$ denotes the expected gain of the evader. 

b. The negative prospect $\mathbb{C}_2^-$. $U_2^- (J)$ represents the loss that makes $J$ less than $J_r$, defined by
\begin{equation}
\label{U2-}
U_2^{-}(J) \triangleq \epsilon_2 \big(-(J-J_r)\big)^{\beta_2}, \;\; J< J_r,
\end{equation}
where $\beta_2 \in (0,1]$ and $\epsilon_2 \in [1,\infty)$. The probability weighting function $w_2^- (p_2)$ is the same as (\ref{probability weighting function2}).

\begin{remark}[\bf{Reduction to the rational case}] In the rational case, the optimal controller is sought through the expected performance index (\ref{probabilistic performance index}). If $\alpha_i = \beta_i=\gamma_i=\epsilon_i=1$ and $w_i^{\pm}(p) = p$, then the prospect terms are identical, and hence the CPT-based performance indices (\ref{CPT performance index}) reduce to $J_1 = J_2 =\mathbb{E}[J]=J_r $. $\hfill \qedsymbol$ \end{remark}

\begin{remark}[\bf{Motivation for the CPT formulation}] The motivation lies in characterizing the subjective evaluation of gain/loss in PE differential games. The rational performance index $J_r$ naturally serves as a reference point, and deviations can be interpreted as gain/loss from $J_r$: for the pursuer, $J<J_r$ corresponds to gain and $J>J_r$ to loss. This makes the gain-loss mechanism of CPT directly compatible with the optimization framework of differential games. $\hfill \qedsymbol$ \end{remark}

By subjectively evaluating the expected deviations from $J_r$, the CPT-based performance indices can be obtained
\begin{equation}
\label{CPT performance index}
J_1= J_r -\mathbb{C}_1(J), \;\; J_2= J_r +\mathbb{C}_2(J)
\end{equation}

Rather than optimizing only $J_r$, the CPT-based formulation optimizes a perceived performance that incorporates irrational perception. The pursuer (resp. the evader) chooses the optimal controller $u^*$ ($v^*$) to minimize $J_1$ (maximize $J_2$). Since $B_i$ is nonsingular, a nonempty class of stabilizing controllers is admitted. The CPT-Nash equilibrium is defined.

\begin{definition} [\bf{CPT-Nash equilibrium}] A pair $\{ u^*,v^* \}$ constitutes a \textit{CPT-Nash equilibrium} for the PE differential game (\ref{dynamic}), (\ref{CPT performance index}) if
\begin{equation}
\label{CPT nash equilibrium}
J_1(u^*, v^*) \leq J_1(u, v^*), \;\; J_2(u^*, v^*) \geq J_2(u^*, v),
\end{equation}
for any admissible controller $u$ and $v$. \hfill$\square$ \end{definition}

\section{CPT-Based PE Capturability and Nash Equilibrium}
In this section, 3.1 gives the main theorem. Then, the conditions in the main theorem are analyzed in 3.2.

\subsection{Main result}
Based on Definitions~1-2 and Lemmas~1-2, the core problem is stated.
\begin{problem} Consider the PE differential games with the irrational perceptions framework (\ref{dynamic}), (\ref{CPT function}) and (\ref{CPT performance index}), the following problem will be investigated.

1) \textit{Capturability}: What are the conditions concerning the system parameters, performance index, and irrationality parameters for achieving successful capture, i.e., $x \rightarrow 0$?

2) \textit{Nash equilibrium}: Can the CPT-Nash equilibrium [i.e., conditions in (\ref{CPT nash equilibrium})] be achieved? \end{problem}

The CPT parameter $\Psi_i$ is first introduced:
\begin{equation}
\label{Psi_i}
\begin{aligned}
\Psi_i \triangleq & (-1)^i \left(\alpha_i \chi_i^{+} \mathcal{M}^{\alpha_i-1} -\epsilon_i \beta_i \chi_i^{-} \mathcal{M}^{\beta_i-1}\right)
\end{aligned}
\end{equation}
where $\chi_i^{+} \!\triangleq \!\int_0^{1/2} \big[-\Phi^{-1}(x)\big]^{\alpha_i} d\big(w_i^{+}(x)\big)$, $\chi_i^{-} \!\triangleq\! \int_0^{1/2} \big[-\Phi^{-1}(x)\big]^{\beta_i} d\big(w_i^{-}(x)\big)$,
and $\mathcal{M} \triangleq x_0^T M x_0$. $\Phi(x)$ is the standard cumulative distribution function. The coefficient $\ell_i$ is defined as
$$
\ell_i \triangleq \begin{cases}\epsilon_i \chi_i^{-} \max \{\beta_i, 1-\beta_i\} h_c^{\beta_i-1} \lambda_{\max }(\Sigma_q), & 0<\beta_i<1 \\ 0, & \beta_i=1 .\end{cases}
$$
and the positive semi-definite matrix $M$ satisfies $x_0^T M x_0 > h_c$ and
\begin{equation}
\label{W}
F^T M +M F = -\Sigma_q
\end{equation}
where
\begin{equation}
\label{hc}
h_{\mathrm c} \triangleq \sqrt{2} \exp\left\{-1-\frac{1}{2}
\int_{0}^{\infty} e^{-\tau}\ln \tau\,d\tau \right\}.
\end{equation}

The following three scenarios will be considered.

\noindent (a) \textbf{Scenario 1.} The pursuer exerts larger control effort, i.e.,
$$
B_1R^{-1}B_1^T>B_2\Pi^{-1}B_2^T.
$$

\noindent (b) \textbf{Scenario 2.} The evader exerts larger control effort, i.e., 
\begin{equation}
\label{condition 21}
B_1R^{-1}B_1^T<B_2\Pi^{-1}B_2^T.
\end{equation}

\noindent (c) \textbf{Scenario 3.} The pursuer and the evader share the same control effort, i.e.,
\begin{equation}
\label{condition 31}
B_1R^{-1}B_1^T = B_2\Pi^{-1}B_2^T.
\end{equation}

The constant matrix $F$ in Scenarios~1-3 is defined as 
\begin{equation}
\label{closed-loop}
F \! \triangleq \!
\begin{cases}
\!(-S_1 E_1 \!+\!S_2 E_2)\mathcal{S}_1^{-1/2}U_1\mathcal{Q}_1^{1/2}, & S_1-S_2>0, \\ \!(-S_1 E_1 \!+\!S_2 E_2)\mathcal{S}_2^{-1/2}U_2(-\mathcal{Q}_2)^{1/2}, & S_1-S_2<0,\\
\!-S^{1/2}(S^{1/2}\mathcal{Q}_1 S^{1/2})^{1/2} S^{-1/2},&  S_1 -S_2=0
\end{cases}
\end{equation}
where $\mathcal{Q}_i \triangleq Q_r+\Psi_i \Sigma_q$, the block-selection matrices $E_1 \triangleq [I_n \; 0] \in\mathbb R^{n\times 2n}$, $E_2 \triangleq [0 \; I_n] \in\mathbb R^{n\times 2n}$, and the specific column-orthonormal matrix $U_i$ satisfies
\begin{equation}
\label{U}
\begin{aligned}
& U_1^T \mathcal{S}_1^{-1/2} \mathcal{S}_2 \mathcal{S}_1^{-1/2} U_1= -\mathcal{Q}_1^{-1/2}\mathcal{Q}_2\mathcal{Q}_1^{-1/2},\\
& U_2^T \mathcal{S}_2^{-1/2} \mathcal{S}_1 \mathcal{S}_2^{-1/2} U_2= (-\mathcal{Q}_2)^{-1/2}\mathcal{Q}_1(-\mathcal{Q}_2)^{-1/2}
\end{aligned}
\end{equation} and $U_i^T U_i = I$. The block matrices $\mathcal{S}_1$, $\mathcal{S}_2$ are defined as
\begin{equation}
\label{blocking matrix}
\mathcal{S}_1 \triangleq
\begin{bmatrix}
S_1 & -S_2\\
-S_2 & S_2
\end{bmatrix}, \;
\mathcal{S}_2 \triangleq
\begin{bmatrix}
S_1 & -S_1\\
-S_1 & S_2
\end{bmatrix}.
\end{equation}

Next, the constants and boundary matrices that will be used in the subsequent theorem are defined as $0 \leq d_1 \leq D_1$, $0 < d_2 < D_2$ for Scenarios~1-2 respectively. Note that all matrix square-root terms involving $D_k, d_k$ are well defined
$$
\begin{aligned}
& \underline{\Gamma}_1 \! \triangleq \!\! \Delta_S^{-1/2} \! \left[\Delta_S^{1/2} \! \left(4\mathcal{Q}_1\!- \!\! D_1^2 \lambda_{\max }\left(S_2\right) \!I\right) \!\Delta_S^{1 / 2}\right]^{1/2} \!\!\Delta_S^{-1/2}, \\
& \overline{\Gamma}_1 \!\triangleq \!\!\Delta_S^{-1/2}\!\left[\Delta_S^{1 / 2}\left(4\mathcal{Q}_1\!-\!\!d_1^2 \lambda_{\min }\left(S_2\right) \!I\right) \!\Delta_S^{1 / 2}\right]^{1/2} \!\!\Delta_S^{-1/2}, \\
& \underline{\Gamma}_2\! \triangleq \!\!\Delta_S^{-1/2}\!\left[\Delta_S^{1 / 2}\left(4\mathcal{Q}_2\!+\!\!d_1^2 \lambda_{\min }\left(S_1\right) \!I\right) \!\Delta_S^{1 / 2}\right]^{1/2} \!\!\Delta_S^{-1/2}, \\
& \overline{\Gamma}_2 \!\triangleq \!\!\Delta_S^{-1/2}\!\left[\Delta_S^{1 / 2}\left(4\mathcal{Q}_2\!+\!\!D_1^2 \lambda_{\max }\left(S_1\right) \!I\right) \!\Delta_S^{1/2}\right]^{1/2} \!\!\Delta_S^{-1/2}
\end{aligned}
$$
and the following two scalar bounds are introduced as
$$
\begin{aligned}
\underline{\Theta}_2 \triangleq &\tfrac12\max\Bigl\{0,\,-\tfrac{D_2}{2}\lambda_{\max}(\Delta_S)\\[-2pt]
&+\Bigl[\tfrac{d_2^2}{4}\lambda_{\min}^2(\Delta_S)+2\lambda_{\min}\bigl(-\Delta_S^{1/2}(\mathcal Q_1+\mathcal Q_2)\Delta_S^{1/2}\bigr)\Bigr]^{1/2}\Bigr\},\\[2pt]
\overline{\Theta}_2 \triangleq &\tfrac12\Bigl\{ \Bigl[\tfrac{D_2^2}{4}\lambda_{\max}^2(\Delta_S) +2\lambda_{\max}\bigl(-\Delta_S^{1/2}(\mathcal Q_1+\mathcal Q_2)\Delta_S^{1/2}\bigr)\Bigr]^{1/2}\\[-2pt]
&-\tfrac{d_2}{2}\lambda_{\min}(\Delta_S)\Bigr\}.
\end{aligned}
$$
where
$$\Delta_S \triangleq \begin{cases}S_1-S_2, \;\; S_1-S_2>0, \\ S_2-S_1, \;\; S_2-S_1>0.\end{cases}$$

For Scenarios~1 and 2, define $\mathcal I_k \triangleq [\underline{y}_k,\overline{y}_k] \subset(h_c,\infty)$, $k\in\{1,2\}$ as a compact admissible interval on which the corresponding conditions are satisfied with common constants $d_k$ and $D_k$, and $\ell_S \triangleq \max_{i\in\{1,2\}} \{ \ell_i\lambda_{\max}(S_i) \}$.

\begin{assumption}
For Scenarios~1-2, the stabilizing Riccati solutions admit a continuous branch on $\mathcal{I}_k$, and the closed-loop trajectory satisfies the endpoint crossing condition.
$\hfill\qedsymbol$
\end{assumption}

The main result is now ready to be presented.
\begin{theorem}[\bf{Capturability and Nash equilibrium}] Under Assumptions~1-2, consider the PE differential games with irrational perceptions framework (\ref{dynamic}), (\ref{CPT function}) and (\ref{CPT performance index}). The capture and CPT-Nash equilibrium (\ref{CPT nash equilibrium}) can be achieved if 

(a) In \textbf{Scenario 1}, there exist constants $0 \leq d_1 \leq D_1$ such that the following conditions (\ref{condition 12-a})-(\ref{condition 13-c}) hold true:
        \begin{gather}
        \label{condition 12-a}
        4\mathcal{Q}_1-D_1^2\lambda_{\max}(S_2)I>0
        \\
        \label{condition 13-a}
        \underline{\Gamma}_1-\overline{\Gamma}_2\geq d_1I
        \\
        \label{condition 13-b}
        \overline{\Gamma}_1-\underline{\Gamma}_2\leq D_1I
        \\
        \label{condition 13-c}
        \mathcal{Q}_2>\frac{\lambda_{\max}^2(\Delta_S)}{\lambda_{\min }^3(\Delta_S)}\left[\frac{D_1}{2} \lambda_{\max}(S_1)+\sqrt{\ell_S}\right]^2 I.
        \end{gather}
        
(b) In \textbf{Scenario 2}, there exist constants $0 < d_2 < D_2$ such that the following conditions (\ref{condition 22-a})-(\ref{condition 23-b}) hold true:
        \begin{gather}
        \label{condition 22-a}
        0<\mathcal{Q}_1<-\mathcal{Q}_2
        \\
        \label{condition 22-b}
        \frac{d_2^2}{4}S_1+\frac{\overline{\Theta}_2^2}{\lambda_{\min}(\Delta_S)}I\leq-\mathcal{Q}_2
        \\
        \label{condition 23-a}
        -\mathcal{Q}_2 \leq \frac{D_2^2}{4}S_1+ \frac{\underline{\Theta}_2^{\,2}}{\lambda_{\max}(\Delta_S)}I.
        \\
        \label{condition 23-b}
        \mathcal{Q}_1-\mathcal{Q}_2> D_2\sqrt{\ell_S} I.
        \end{gather}

(c) In \textbf{Scenario 3}, conditions (\ref{condition 32})-(\ref{condition 33}) hold true:
        \begin{gather}
        \label{condition 32}
        \mathcal{Q}_1=-\mathcal{Q}_2>0
        \\
        \label{condition 33}
        \mathcal{Q}_1 \geq \max \left\{\ell_1, \ell_2\right\} I
        \end{gather}
\end{theorem}

\begin{IEEEproof} The proof is deferred to Section~4. $\hfill \blacksquare$\end{IEEEproof} 

\subsection{Analysis of Conditions}
The conditions in Theorem~1 are discussed in this subsection.

\begin{proposition} Consider the CPT-function (\ref{CPT function}) with gain sensitivity $\alpha_i$, loss sensitivity $\beta_i$, loss multiplier $\epsilon_i$ and fixed $\mathcal{M} > h_c$, the irrationality parameter $\Psi_1$ is strictly increasing in $\beta_1$, $\epsilon_1$ and strictly decreasing in $\alpha_1$, while $\Psi_2$ is strictly decreasing in $\beta_2$, $\epsilon_2$ and strictly increasing in $\alpha_2$.\end{proposition}

\begin{IEEEproof} See Appendix A. $\hfill \blacksquare$ \end{IEEEproof} 

Based on Proposition 1, how these irrationality parameters affect the conditions in Theorem 1 is examined.

(a) \textbf{Scenario 1.} Under condition (\ref{lemma}), capture is achievable in the rational case. Condition (\ref{condition 12-a}), with the lower bound on $\mathcal{Q}_2$ in (\ref{condition 13-c}) guarantees that the fixed-point mapping is well-defined and continuous. Conditions (\ref{condition 13-a})-(\ref{condition 13-b}) impose lower and upper admissible bounds on the irrationality parameters. In addition, condition (\ref{condition 13-c}) provides a stability bound characterized by $\ell_i$. Consequently, the corresponding Riccati solutions and capturability are established and the CPT-Nash equilibrium is verified. Specifically,

For the \textbf{Pursuer}, $\Psi_1$ increases as the \textbf{gain sensitivity $\alpha_1$} decreases, the \textbf{loss sensitivity $\beta_1$} increases and the \textbf{loss multiplier $\epsilon_1$} increases. It follows that \textit{if the pursuer exerts larger control effort, then smaller gain sensitivity, larger loss sensitivity and larger loss multiplier of the irrational pursuer on its performance index facilitate capture.}

For the \textbf{Evader}, larger $\alpha_2$ and smaller $\beta_2$ and $\epsilon_2$ increase $\Psi_2$ and hence facilitating the satisfaction of (\ref{condition 13-c}) and capture.

(b) \textbf{Scenario 2.} Under condition (\ref{condition 21}), capture is not achievable for the rational case, but can be achieved under conditions (\ref{condition 22-a})-(\ref{condition 23-b}). The effects of the CPT parameters on capturability are opposite to those in Scenario~1.

(c) \textbf{Scenario 3.}
Condition (\ref{condition 32}) requires a balance between $Q_r$ and $(\Psi_1+\Psi_2)\Sigma_q$ for the coupled CPT-based Riccati equations to admit a common solution.

The consistency of the proposed CPT framework with rational PE differential games is then examined.

\textbf{Rational Case.} When all irrational factors degrade to the rational case, i.e., the loss multiplier parameter $\epsilon_i=1$, sensitivity parameter $\alpha_i = \beta_i =1$, the perceptual distortion of probabilities $\gamma_i =1$ and the probability sensitivity function $w_i^{+}(x) =w_i^{-}(x)=x$, one gives that $J_1=J_2=J_r$, $\Psi_1=\Psi_2=0$, $\ell_1=\ell_2=0$.

(a) \textbf{Scenario 1.}
The lower and upper bounds are set to $d_1 = D_1 = 0$, then the conditions (\ref{condition 12-a})  and (\ref{condition 13-c}) necessarily turn into the same condition $\mathcal{Q}_1 = \mathcal{Q}_2 = Q_r > 0$. It is easily observed that conditions (\ref{condition 13-a})-(\ref{condition 13-b}) hold true, then recalling Lemma 2, the result can be derived from (\ref{lemma}).

(b) \textbf{Scenario 2.} In the rational case, $\Psi_1=\Psi_2=0$, and hence $\mathcal Q_1+\mathcal Q_2=2Q_r>0$, which violates condition~(\ref{condition 22-a}). Therefore, the conditions of Scenario~2 cannot hold in the rational case, which remains consistent with Lemma 2.

(c) \textbf{Scenario 3.} In this case $Q_r$ is required to be $0_{3\times3}$, which contradicts $Q_r > 0$, and the result can be derived.

\section{Ancillary results and proof of main result}
Section~4.1 reformulates the CPT-based performance index, Section~4.2 analyzes the solvability of the CPT-based AREs and capturability. Section~4.3 provides a feasible fixed-point computation method.

\subsection{An equivalent problem for CPT-based optimal controller and AREs}
In this subsection, the CPT-based performance index is equivalently reformulated to separate the variance dependent and constant terms.
\begin{proposition} Consider the CPT-based performance index (\ref{CPT performance index}) subject to (\ref{dynamic}),(\ref{nominal performance index}), and prospect function (\ref{CPT function}). For any given prospect parameters $\alpha_i,\beta_i, \gamma_i, \epsilon_i$ and $w_i^{+,-}$, the equivalent form of (\ref{CPT performance index}) is introduced as follows:
\begin{equation}
\label{proposition 1}
J_i=J_r+(-1)^i (\chi_i^+\sigma^{\alpha_i}-\epsilon_i\chi_i^-\sigma^{\beta_i}).
\end{equation}
\end{proposition}

\begin{IEEEproof} See Appendix B. $\hfill \blacksquare$ \end{IEEEproof} 

Then a suitable coordinate transformation is performed to obtain an extended form of the associated CPT-based coupled AREs. Define the state-dependent term $y(t)\triangleq \int_0^t x(\tau)^T \Sigma_q x(\tau) d \tau$, $y(0)=0$, for any admissible controllers, the accumulated value is finite and is denoted by $y_{\infty} \triangleq \lim_{t\rightarrow\infty}y(t) = \int_{0}^{\infty}x^{T}(t)\Sigma_qx(t)dt$. The system dynamic (\ref{dynamic}) can be extended to
$$
\left\{\begin{array}{l}
\dot{x}=B_1 u-B_2 v \\
\dot{y}=x^T \Sigma_q x.
\end{array}\right.
$$

To derive the optimal controllers concerning the CPT performance indices, $\lambda_i$ and $\mu_i$ are denoted as the costates associated with $x$ and $y$, respectively. The Hamiltonian for pursuer (resp. evader) can be expressed as:
$$
\begin{aligned}
H_i \!= \!& x^T Q_r x+u^T R u-v^T \Pi v+\lambda_i^T\left(B_1 u-B_2 v\right)+\mu_i x^T \Sigma_qx.
\end{aligned}
$$

For a finite truncation horizon $T$, the terminal condition associated with $y$ is $\mu_i(T) =(-1)^i [\alpha_i\chi_i^+y(T)^{\alpha_i-1}-\epsilon_i\beta_i\chi_i^-y(T)^{\beta_i-1}]$. By applying the relation $\dot{\mu}_i=-\partial H_i/\partial y=0$, one implies $\mu_i(t)\equiv\mu_i(T)$ on $[0,T]$. For $y(T)\rightarrow y_\infty$ as $T\rightarrow\infty$, at the solution $y_\infty=\mathcal M=x_0^TMx_0$, oen can observe $\mu_i(t)\equiv\Psi_i$. Then considering $\frac{\partial H_1}{\partial u}=0$ and $\frac{\partial H_2}{\partial v}=0$, the optimal controllers can be expressed as
\begin{equation}
\label{CPT controller}
u^* = -R^{-1} B_1^T P_1 x, \;\; v^* = -\Pi^{-1} B_2^T P_2 x.
\end{equation}

For a fixed $\Psi_i$, the associated value function admits the quadratic form $x^T P_i x$. Substituting (\ref{CPT controller}) into the costate equations yields the coupled CPT-based AREs
\begin{equation}
\label{CPT Riccati equation}
\begin{aligned}
0=& \mathcal{Q}_1- P_1 S_1 P_1-P_2 S_2 P_2+P_1 S_2 P_2 +P_2 S_2 P_1 \\
0=& \mathcal{Q}_2 +P_1 S_1 P_1+P_2 S_2 P_2-P_2 S_1 P_1 -P_1 S_1 P_2.
\end{aligned}
\end{equation}
For the subsequent analysis, $\Psi_i(y)$ and $\mathcal{Q}_i(y)$ are evaluated at $\mathcal{M}=y>h_c$ as $\Psi_i$ and $\mathcal{Q}_i$, respectively.
\subsection{Solvability of CPT-based AREs and Capturability Analysis}
This subsection introduces the solvability of CPT-based AREs.

\begin{proposition} The CPT-based PE differential games (\ref{dynamic}), (\ref{CPT performance index}) admit symmetric positive-definite solutions $P_1$ and $P_2$ in each Scenario, if the corresponding conditions hold true: specifically, conditions (\ref{lemma}), (\ref{condition 12-a})-(\ref{condition 13-c}) for Scenario 1; (\ref{condition 21}), (\ref{condition 22-a})-(\ref{condition 23-a}) for Scenario~2; and (\ref{condition 31}) and (\ref{condition 32}) for Scenario~3. \end{proposition}

\begin{IEEEproof} See Appendix C. $\hfill \blacksquare$ \end{IEEEproof}

Then the capturability characterizations are derived. Under the optimal controllers (\ref{CPT controller}), it suffices to show that the closed-loop system $F$ is Hurwitz, and the three scenarios are considered separately.

(a) \textbf{Scenario 1.} From the fixed-point construction, one obtains $0\leq W\leq D_1I$, then the closed-loop matrix can be expressed as $F=-\Delta_S P_2-\frac{1}{2} S_1 W$. Applying the similarity transformation of $F$, it follows that
\begin{equation}
\label{F similarity}
\tilde{F}=-\Delta_S^{1/2} P_2 \Delta_S^{1/2}-\frac{1}{2} \Delta_S^{-1/2} S_1 W \Delta_S^{1/2}.
\end{equation}

Along with (\ref{eq:XWYW}), one yields $\frac{\tilde{F}+\tilde{F}^T}{2} \leq - f I$, where $f \triangleq \sqrt{\lambda_{\min}(\Delta_S) \lambda_{\min}(\mathcal{Q}_2)}-\frac{D_1}{2} \lambda_{\max}(S_1) \sqrt{\frac{\lambda_{\max}(\Delta_S)}{\lambda_{\min}(\Delta_S)}}$. Condition (\ref{condition 13-c}) guarantees that $\tilde{F}$ is Hurwitz, and thus $F$ is Hurwitz, which establishes capturability in Scenario 1.

\textbf{Scenarios 2 and 3.} From Appendix~C, $P_1-P_2>0$ in both scenarios, subtracting the two equations in (\ref{CPT Riccati equation}) gives
$$
F^T (P_1-P_2)+(P_1-P_2) F=-(\mathcal{Q}_1-\mathcal{Q}_2).
$$
For Scenario~2, condition~(\ref{condition 23-b}) yields $\mathcal{Q}_1-\mathcal{Q}_2>0$, whereas for Scenario~3, condition (\ref{condition 32}) gives $\mathcal{Q}_1-\mathcal{Q}_2 =2\mathcal{Q}_1>0$. Hence, $F^T (P_1-P_2)+(P_1-P_2) F<0$ in both cases.

\subsection{CPT-Nash Equilibrium and Solution Characterization}
In this subsection, the CPT-Nash equilibrium is verified, and the Riccati solutions are parameterized.

For Scenarios~1 and 2, $(P_{1,k}(y),P_{2,k}(y))$ is defined on $\mathcal{I}_k$, then the closed-loop matrix can be defined as $F_k(y) = -S_1P_{1,k}(y)+S_2P_{2,k}(y)$. Since $F_k(y)$ is Hurwitz, the Lyapunov equation
$$
F_k^T(y)M_k(y)+M_k(y)F_k(y)=-\Sigma_q
$$
admits a unique solution $M_k(y)>0$. Since the Riccati branch $\big(P_{1,k}(y),P_{2,k}(y)\big)$ is continuous on $\mathcal I_k$, the accumulated term $\hat{Y}_k(y) = x_0^T M_k(y)x_0$ is also continuous. This implies that the CPT coefficient $\Psi_i(y)$ is self-consistent with the closed-loop trajectory if $y=\hat{Y}_k(y)$.

To establish the existence of the solution, denote $K_-^{(k)}>0$ as an upper bound satisfying $\|F_k(\underline{y}_k)\|\leq K_-^{(k)}$, and the closed-loop matrix satisfies $\|e^{F_k(\bar{y}_k)t}\| \leq C_+^{(k)}e^{-a_+^{(k)}t}$. The constants $K_-^{(k)}$, $C_+^{(k)}$, and $a_+^{(k)}$ are explicitly bounded. 

\textbf{Scenario 1.} Appendix C yields the lower endpoint bound
$$
K_-^{(1)}\!\! =\!\!\frac{\lambda_{\max}(\Delta_S)}{2}\!\sqrt{\!\frac{ D_1^2\lambda_{\max}(S_1) \!+\!4\lambda_{\max}\big(\mathcal{Q}_2(\underline{y}_1)\big)}{\lambda_{\min}(\Delta_S)}} \!+\!\frac{D_1}{2}\lambda_{\max}(S_1).
$$
On the other hand, one may choose $C_+^{(1)}=\sqrt{\frac{\lambda_{\max}(\Delta_S)} {\lambda_{\min}(\Delta_S)}}$ and
$$
\begin{aligned}
a_{+}^{(1)}=& \sqrt{\lambda_{\min}(\Delta_S) \lambda_{\min}(\mathcal{Q}_2(\overline{y}_1))}-\frac{D_1}{2} \lambda_{\max}(S_1) \sqrt{\frac{\lambda_{\max }(\Delta_S)}{\lambda_{\min }(\Delta_S)}}.
\end{aligned}
$$
The convergence rate $a_+^{(1)}>0$ is guaranteed by condition~(\ref{condition 13-c}) .

\textbf{Scenario 2.} The coupled AREs imply that
$$
K_-^{(2)} \!=\! \frac{D_2}{2}\lambda_{\max}(S_1) \sqrt{\!\frac{\lambda_{\max}(S_2)} {\lambda_{\min}(\Delta_S)}} \!+\!\lambda_{\max}(S_2) \sqrt{\!\frac{\lambda_{\max}(-\mathcal{Q}_2(\underline y_2))} {\lambda_{\min}(\Delta_S)}}.
$$

Invoking $d_2I \leq W \leq D_2I$, one yields that $C_+^{(2)} = \sqrt{\frac{D_2}{d_2}}$ and $a_+^{(2)} = \frac{ \lambda_{\min} \big(\mathcal{Q}_1(\overline{y}_2) -\mathcal{Q}_2(\overline{y}_2)\big)}{D_2}$ with $a_+^{(2)}>0$. In Scenarios~1-2, the endpoint conditions in Assumption~2 are $\frac{ \lambda_{\min}(\Sigma_q)\|x_0\|^2}{2K_-^{(k)}} \geq \underline{y}_k$ and $\frac{(C_+^{(k)})^2 \lambda_{\max}(\Sigma_q)\|x_0\|^2}{2a_+^{(k)}} \leq \overline{y}_k$.

According to the continuity of $\widehat Y_k(y)$, the intermediate value theorem guarantees that at least one $y_k^*\in\mathcal I_k$ satisfying $y_k^*=\hat{Y}_k(y_k^*)$ exists. In Scenario~3, self-consistency is imposed through the existence of $M>0$ satisfying (\ref{W}) with $\mathcal M=x_0^TMx_0>h_c$ in Theorem~1. Then the CPT-Nash equilibrium is established in the following proposition.

\begin{proposition} Consider the CPT-based PE differential games (\ref{dynamic}) and (\ref{CPT performance index}) with the optimal controllers (\ref{CPT controller}) satisfying (\ref{CPT Riccati equation}). Under the corresponding conditions of Theorem~1, the CPT-Nash equilibrium (\ref{CPT nash equilibrium}) is achieved. \end{proposition}

\begin{IEEEproof} See Appendix D. $\hfill \blacksquare$ \end{IEEEproof} 

The Riccati solutions can be parameterized as follows.

\begin{corollary} Consider the stabilizing solutions of equations (\ref{CPT Riccati equation}) under the corresponding conditions of Proposition~3. The following parameterizations hold in Scenarios~1-3.

(a) \textbf{Scenario 1.}
\begin{equation}
\label{Solution 1}
P_1 = E_1\mathcal{S}_1^{-1/2} U_1\mathcal{Q}_1^{1/2}, \; P_2 = E_2\mathcal{S}_1^{-1/2} U_1\mathcal{Q}_1^{1/2}.
\end{equation}

(b) \textbf{Scenario 2.}
\begin{equation}
\label{Solution 2}
P_1 \!=\! E_1\mathcal{S}_2^{-1/2} U_2(\!-\mathcal{Q}_2)^{1/2}, P_2 \!=\! E_2\mathcal{S}_2^{-1/2} U_2(\!-\mathcal{Q}_2)^{1/2}.
\end{equation}

(c) \textbf{Scenario 3.}
\begin{equation}
\label{Solution 3}
P_1=C+S^{-1/2} (S^{1/2}\mathcal{Q}_1S^{1/2})^{1/2}
S^{-1/2}, P_2=C
\end{equation}
where $C$ is any matrix satisfying $C=C^T>0$. \end{corollary}

\begin{IEEEproof} See Appendix E. $\hfill \blacksquare$ \end{IEEEproof} 

\section{Illustrative Examples}
In this section, numerical examples are presented to illustrate the effectiveness of the proposed approach. All CPT parameters are fixed at their rational values, i.e., $\alpha_1=\alpha_2=\beta_1=\beta_2=\epsilon_1=\epsilon_2=\gamma_1=\gamma_2=1$, with only the parameter under investigation varied.

\subsection{Scenario 1}
The system matrices and initial states are selected as $B_1=B_2=I_{3\times3}$. The initial locations of the pursuer and evader are (0, 10, 0) and (10, 15, 5). The weighting matrices in the performance index (\ref{probabilistic performance index}) with $Q_r=I_{3 \times 3}$, $R= 0.9 I_{3 \times 3}$, $\Pi=I_{3 \times 3}$ are considered, and $S_1-S_2=\frac{1}{9}I_3>0$. The uncertainty structure matrix is selected as $Q_\Delta=I_{3\times3}$, the uncertainty scale parameter is $q=0.9$.

\begin{figure}[htbp]
    \centering
    \includegraphics[width=\linewidth]{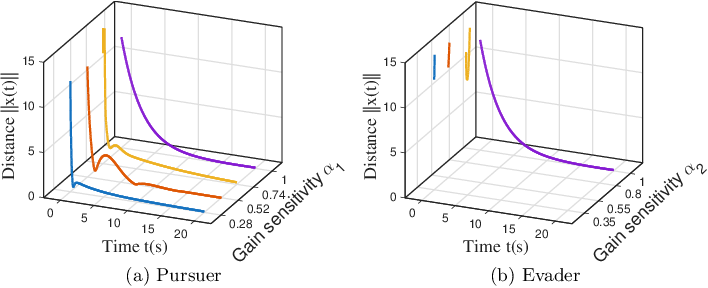}
    \caption{Scenario 1: Gain sensitivity}
    \label{s1a1}
\end{figure}

\textbf{Gain sensitivity $\alpha_i$ :} How varying gain sensitivity $\alpha_i$ affects the PE dynamics is first examined. Fig.~\ref{s1a1} shows that successful capture occurs for all $\alpha_1$, which satisfies conditions~(\ref{condition 12-a})-(\ref{condition 13-c}). In addition, capture can only be achieved in the rational case $\alpha_2=1$, which satisfies conditions~(\ref{condition 12-a})-(\ref{condition 13-c}).

\begin{figure}[htbp]
    \centering
    \includegraphics[width=\linewidth]{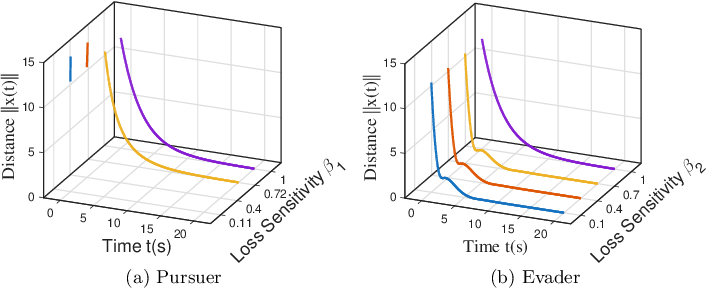}
    \caption{Scenario 1: Loss sensitivity}
    \label{s1b1}
\end{figure}

\textbf{Loss sensitivity $\beta_i$:}
Next, the effect of loss sensitivity $\beta_i$ is studied in Fig.~\ref{s1b1}, where $\beta_1 = 1$ satisfies conditions~(\ref{condition 12-a})-(\ref{condition 13-c}). As expected, the larger $\beta_1$ leads to a successful capture. By contrast, with the pursuer's loss sensitivity $\beta_1 = 1$ fixed, all capture can be achieved with the evader's loss sensitivity.

\begin{figure}[htbp]
    \centering
    \includegraphics[width=\linewidth]{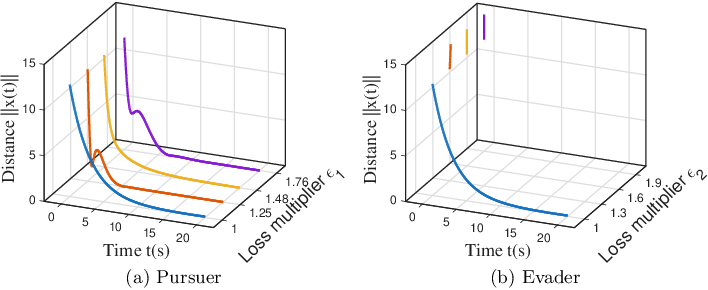}
    \caption{Scenario 1: Loss multiplier}
    \label{s1r1}
\end{figure}

\textbf{Loss multiplier $\epsilon_i$:} 
The effect of the loss multiplier $\epsilon_i$ is examined in Fig.~\ref{s1r1}. All the $\epsilon_1$ satisfy the conditions~(\ref{condition 12-a})-(\ref{condition 13-c}). Capture is no longer achieved once the evader's loss multiplier $\epsilon_2$ increases beyond a certain threshold.

\subsection{Scenario 2}
In this case, the weighting matrices in the performance index (\ref{probabilistic performance index}) with $Q_r= 0.2I_{3 \times 3}$, $R= I_{3 \times 3}, \Pi=0.9 I_{3 \times 3}$, $q=1.03$ are considered. The other settings are the same as those in Scenario 1. 

\begin{figure}[htbp]
    \centering
    \includegraphics[width=\linewidth]{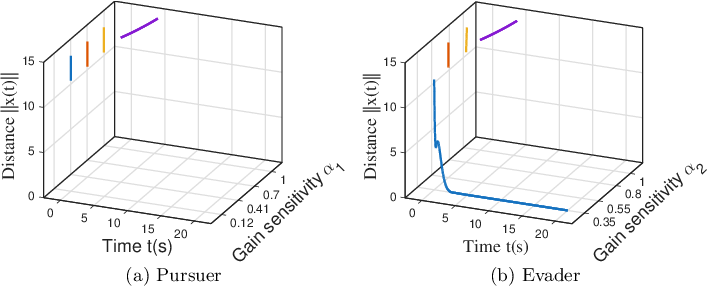}
    \caption{Scenario 2: Gain sensitivity}
    \label{s2a1}
\end{figure}

\textbf{Gain sensitivity $\alpha_i$ :} Fig.~\ref{s2a1} shows that successful capture cannot be achieved in the rational case, and can be achieved in the irrational case $\alpha_2=0.35$, which satisfies conditions~(\ref{condition 22-a})-(\ref{condition 23-b}).

\begin{figure}[htbp]
    \centering
    \includegraphics[width=\linewidth]{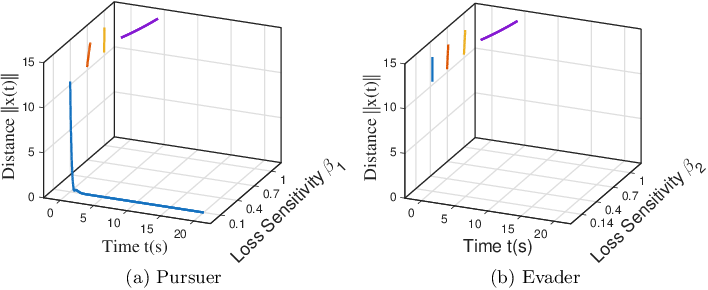}
    \caption{Scenario 2: Loss sensitivity}
    \label{s2b1}
\end{figure}

\textbf{Loss sensitivity $\beta_i$:}
The simulation is shown in Fig.~\ref{s2b1}. As expected, the smaller $\beta_1$ leads to a successful capture. On the other hand, capture cannot be achieved with the evader's loss sensitivity $\beta_2 \in \{ 0.14,0.4, 0.7, 1\}$. 

\begin{figure}[htbp]
    \centering
    \includegraphics[width=\linewidth]{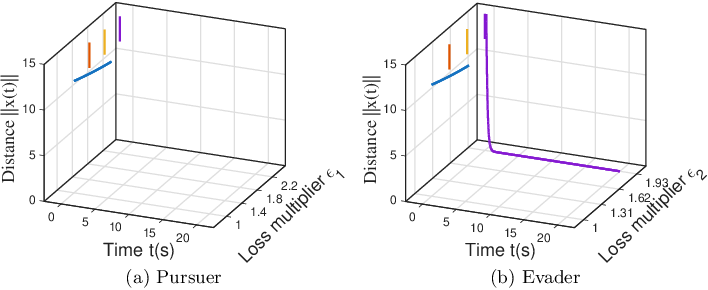}
    \caption{Scenario 2: Loss multiplier}
    \label{s2r1}
\end{figure}

\textbf{Loss multiplier $\epsilon_i$:} 
The capture results are reported in Fig.~\ref{s2r1}. Capture can be achieved once the evader's loss multiplier $\epsilon_2$ increases beyond a certain threshold. 

\section{Conclusion}
To characterize the subjective bias in perceiving the probabilistic characteristics of environmental uncertainty in practical human-machine systems, this paper develops a CPT-based PE differential game framework to model irrational behaviors. By bridging irrational behavioral theory with game-based control, sufficient conditions for capturability and the CPT-Nash equilibrium are derived. Future work will extend the proposed framework to networked multi-player differential games.

\appendices

\section{Proof of Proposition 1}
Consider the CPT function (\ref{CPT function}), define the weighted geometric means
$$
\begin{aligned}
& G_i^+(\alpha_i) \triangleq \exp \left\{ {\frac{\int_0^{1/2} a(p)^{\alpha_i}\ln a(p)d w_i^+(p)}{\chi_i^{+}}} \right\},\\
& G_i^-(\beta_i) \triangleq \exp \left\{ {\frac{ \int_0^{1/2} a(p)^{\beta_i}\ln a(p)d w_i^-(p)}{\chi_i^{-}}} \right\}.
\end{aligned}
$$
where $\ln(\cdot)$ denotes the natural logarithm to base $e$, and $a(p)\triangleq-\Phi^{-1}(p)$. Differentiating (\ref{Psi_i}) with respect to
$\alpha_i$ and $\beta_i$ delivers that
$$
\begin{aligned}
\frac{\partial \Psi_i}{\partial \alpha_i}
=&(-1)^i \mathcal{M}^{\alpha_i-1}\chi_i^{+}\Big[1+\alpha_i \ln \big(\mathcal{M} G_i^+(\alpha_i) \big)\Big],\\
\frac{\partial \Psi_i}{\partial \beta_i}
=&(-1)^{i+1} \epsilon_i \mathcal{M}^{\beta_i-1}\chi_i^{-}\Big[1+\beta_i \ln \big(\mathcal{M} G_i^-(\beta_i) \big)\Big]
\end{aligned}
$$

For the weighted geometric means $G_i^\pm(\cdot)$, one implies that $G_i^\pm(s)\ge e^{-1}/h_c$. For any $s \in (0,1]$, $e^{-1/s} \leq e^{-1}$, and $\frac{e^{-1/s}}{G_i^\pm(\cdot)} \leq h_c$, hence, the condition $\mathcal{M}>h_c$ implies that $\mathcal{M}> \frac{e^{-1/s}}{G_i^\pm(s)}$, and further $\mathcal{M}G_i^+(\alpha_i)>e^{-1/\alpha_i}$, $\mathcal{M}G_i^-(\beta_i)>e^{-1/\beta_i}$ with $\alpha_i, \beta_i \in(0,1]$.

Then the monotonicity in $\epsilon_i$ is considered. Recalling the representation (\ref{Psi_i}), with the fixed $(\alpha_i,\beta_i)$, differentiating $\Psi_i$ with respect to $\epsilon_i$ yields $\frac{\partial \Psi_i}{\partial \epsilon_i}= (-1)^{i+1}\,\beta_i \chi_i^{-}\mathcal{M}^{\beta_i-1}$.

Since $\beta_i>0$, $\chi_i^->0$ and $\mathcal{M}^{\beta_i-1}>0$, one obtains $\frac{\partial \Psi_1}{\partial \epsilon_1} > 0$ and $\frac{\partial \Psi_2}{\partial \epsilon_2}< 0$. Hence, $\Psi_1$ is strictly increasing in $\epsilon_1$ and $\Psi_2$ is strictly decreasing in $\epsilon_2$.
 
\section{Proof of Proposition 2}
Recalling (\ref{mathematical expectation}) and (\ref{Var J}), the cumulative distribution function is symmetric about $1/2$. This implies that for the pursuer, the gain corresponds to $J \leq J_r$, whereas for the evader it corresponds to $J \geq J_r$. Hence, for any $h>0$, the corresponding probability of the positive prospect follows
$$
\begin{aligned}
p_i\left\{U_i^{+}(J)>h\right\} & =p_i \left\{\left[(-1)^i\left(J-J_r\right)\right]^{\alpha_i}>h\right\} \\
& =1-\Phi\left(\frac{h^{1 / \alpha_i}}{\sigma}\right).
\end{aligned}
$$

It follows from (\ref{CPT function}) that
\begin{equation}
\label{CPT trans2}
\mathbb{C}_i^+(J)=\int_0^{\infty} w_i^{+}\left[1-\Phi\left(\frac{h^{1 / \alpha_i}}{\sigma}\right)\right] d h .
\end{equation}

Define $z \triangleq \frac{h^{1/\alpha_i}}{\sigma}$, one implies that $dh = \alpha_i \sigma^{\alpha_i}z^{\alpha_i-1} dz$, then recalling (\ref{CPT trans2}), one can observe the result
\begin{equation}
\label{pre C+}
\begin{aligned}
\mathbb{C}_i^+(J) \!\!=& \int_0^{\infty} w_i^{+}(1-\Phi(z)) \alpha_i \sigma^{\alpha_i} z^{\alpha_i-1} d z \\
= & \!\sigma^{\alpha_i}\!\left[z^{\alpha_i} w_i^{+}(1\!-\!\Phi(z))\right]_0^{\infty} \!-\!\sigma^{\alpha_i} \int_0^{\infty}\!\!\! z^{\alpha_i} d\left[w_i^{+}(1\!-\!\Phi(z))\right]
\end{aligned}
\end{equation}

Since the boundary term in (\ref{pre C+}) vanishes due to $\alpha_i>0$ and the boundedness of $w_i^+(1/2)$, one yields the result
\begin{equation}
\label{C+}
\begin{aligned}
\mathbb{C}_i^+(J) =& -\sigma^{\alpha_i} \int_{1 / 2}^0\left[\Phi^{-1}(1-x)\right]^{\alpha_i} d\left(w_i^{+}(x)\right) \\
=& \chi_i^{+} \sigma^{\alpha_i}
\end{aligned}
\end{equation}

Similarly, the corresponding probability of the negative prospect is obtained as
$$
\begin{aligned}
p_i\left\{U_i^{-}(J)>h\right\} & =p_i \left\{(-1)^{i+1} \eta>\frac{\left(h / \epsilon_i\right)^{1 / \beta_i}}{\sigma}\right\} \\
& =1-\Phi\left(\frac{\left(h / \epsilon_i\right)^{1 / \beta_i}}{\sigma}\right),
\end{aligned}
$$
and the expected negative prospect yields that $\mathbb{C}_i^-(J) = \epsilon_i \chi_i^{-} \sigma^{\beta_i}$. Substituting (\ref{C+}) and the expected negative prospect into (\ref{CPT function}) leads directly to (\ref{proposition 1}). The proof is complete.

\section{Proof of Proposition 3} (a) \textbf{Scenario 1.}
As a first step, set $X \triangleq 2P_1$, $Y \triangleq 2P_2$, $W \triangleq X-Y=2(P_1-P_2)$ and $\Delta_S=S_1-S_2>0$, and the original equation (\ref{CPT Riccati equation}) is equivalent to
\begin{equation}
\label{trans riccati equation}
\begin{aligned}
4 \mathcal{Q}_1 &= X\Delta_S X + W S_2 W,\\
-4 \mathcal{Q}_2 &= W S_1 W - Y\Delta_S Y.
\end{aligned}
\end{equation}

On the nonempty compact convex set $\mathcal{K}_1 \triangleq \{W\in\mathbb S^n: d_1 I \leq W \leq D_1 I\}$, for each $W\in \mathcal{K}_1$, the fixed-point map is constructed as
\begin{equation}
\label{eq:XWYW}
\begin{aligned}
X(W) \!\triangleq \! & \Delta_S^{-1/2} \Big[ \Delta_S^{1/2} \big(
4\mathcal{Q}_1\!-\!WS_2W \big) \Delta_S^{1/2} \Big]^{1/2} \Delta_S^{-1/2},\\
Y(W) \!\triangleq \! & \Delta_S^{-1/2} \Big[ \Delta_S^{1/2} \big( WS_1W\!+\!4\mathcal{Q}_2\big) \Delta_S^{1/2} \Big]^{1/2} \Delta_S^{-1/2}
\end{aligned}
\end{equation}
and
\begin{equation}
\label{fixed-point map}
T(W) \triangleq X(W)-Y(W).
\end{equation}

Based on the condition (\ref{condition 12-a}) and condition (\ref{condition 13-c}), the positive definiteness in (\ref{eq:XWYW}) is guaranteed. For any $W\in \mathcal{K}_1$ and any $\xi\in\mathbb R^n$, one delivers that
$$
d_1^2\lambda_{\min}(S_i)I \leq WS_iW \leq D_1^2\lambda_{\max}(S_i)I.
$$
Then one can deduce that $X(\cdot)$, $Y(\cdot)$, and $T(\cdot)$ are well-defined and continuous on $\mathcal K_1$. Applying the operator monotonicity and conditions (\ref{condition 13-a})-(\ref{condition 13-b}) implies that
$$
d_1I \leq \underline{\Gamma}_1-\overline{\Gamma}_2 \leq T(W)\leq \overline{\Gamma}_1-\underline{\Gamma}_2 \leq D_1I.
$$
Thus, one $W^\ast\in\mathcal K_1$ satisfying $T(W^\ast)=W^\ast$ is guaranteed by Brouwer's fixed-point theorem. Invoking the initial transformation $X =2P_1$ and $Y=2P_2$, it follows that
$$
P_1 = \frac{1}{2}X^*>0, \;\; P_2 \triangleq \frac{1}{2}Y^*>0.
$$
Up to this point, under conditions (\ref{lemma}), (\ref{condition 12-a})-(\ref{condition 13-c}), the existence of solutions $P_1$ and $P_2$ to the CPT-Riccati equations (\ref{CPT Riccati equation}) can be guaranteed in Scenario 1.

(b) \textbf{Scenario 2.} Adding the two coupled Riccati
equations in (\ref{CPT Riccati equation}) gives
$$
P_1\Delta_S P_2+P_2\Delta_S P_1=-(\mathcal{Q}_1+\mathcal{Q}_2).
$$
For each $W\in\mathcal K_2$ where $\mathcal{K}_2 \! \triangleq \! \{ W\in\mathbb S^n: d_2I\leq W\leq D_2I \}$, define
$$ U(W) \!\triangleq \! \frac{1}{2} \left[\left(V^2(W)\!-\!2\Delta_S^{1 / 2}(\mathcal{Q}_1\!+\!\mathcal{Q}_2) \Delta_S^{1 / 2}\!\right)^{1/2}\!\!-\!V(W)\right],
$$
and one can obtain
$$
P_2(W) \triangleq \Delta_S^{-1/2}U(W)\Delta_S^{-1/2}
$$
where $V(W) \triangleq \frac{1}{2}\Delta_S^{1/2}W\Delta_S^{1/2}$. Since $V(W)>0$, the operator monotonicity of the principal square root gives $U(W)>0$, and $P_2(W)>0$. Moreover, with $P_1(W)=P_2(W)+\frac{1}{2}W$, the preceding Riccati relation is satisfied. By the definitions of $\underline{\Theta}_2$ and $\overline{\Theta}_2$, one obtains
$$
\frac{\underline{\Theta}^2_2}{\lambda_{\max}(\Delta_S)}I
\leq
P_2(W)\Delta_SP_2(W)
\leq
\frac{\overline{\Theta}^2_2}{\lambda_{\min}(\Delta_S)}I.
$$
Hence, conditions~(\ref{condition 22-b}) and (\ref{condition 23-a}) imply that
$$
\frac{d_2^2}{4}S_1 \leq -\mathcal{Q}_2-P_2(W)\Delta_SP_2(W) \leq \frac{D_2^2}{4}S_1.
$$
Define the continuous mapping
$$
T_2(W) \!\!\triangleq \!2S_1^{-1/2} \!\left[\!S_1^{1/2} \!\left(-\mathcal{Q}_2\!\!-\!\!P_2(W)\Delta_SP_2(W) \right) \! S_1^{1/2} \! \right]^{1/2} \!\! S_1^{-1/2}.
$$
Then the preceding bounds yield $d_2I\leq T_2(W)\leq D_2I$, and one delivers $T_2(\mathcal{K}_2) \subseteq \mathcal{K}_2$. By the same argument as in Scenario~1, there exists $W^*\in\mathcal{K}_2$ satisfying $T_2(W^*)=W^*$, which implies that $P_1(W^*)$ and $P_2(W^*)$ are positive-definite solutions of (\ref{CPT Riccati equation}). This completes the proof for Scenario~2.

(c) \textbf{Scenario 3.} Adding the two equations in (\ref{CPT Riccati equation}) yields
\begin{equation}
\label{scenario 3}
2 Q_r+(\Psi_1+\Psi_2) \Sigma_q=0.
\end{equation}

Consequently, (\ref{scenario 3}) is the compatibility condition in Scenario 3, and the coupled Riccati equations (\ref{CPT Riccati equation}) reduce to
\begin{equation}
\label{ARE3}
(P_1-P_2) S_1 (P_1-P_2)=\mathcal{Q}_1
\end{equation}
and its solution $P_1 -P_2$ can be easily given by $P_1-P_2 = S_1^{-1/2} \left[ S_1^{1/2} \mathcal{Q}_1 S_1^{1/2} \right]^{1/2} S_1^{-1/2}>0$. The reduced equation admits solutions $P_1>0$ and $P_2>0$.

At this point, the existence of symmetric positive-definite solutions to the coupled AREs (\ref{CPT Riccati equation}) is established.

\section{Proof of Proposition 4} The optimal controllers $u^*$ and $v^*$ are evaluated along the deviated trajectory $x_1$. For the pursuer, when the evader adopts the optimal controller $v^*$, the deviation of the performance index can be obtained as
$$
\begin{aligned}
J_1\left(u, v^*\right)-& J_1\left(u^*, v^*\right) = \int_0^{\infty} (u-u^*)^T R (u-u^*) dt \\
& +\chi_1^{+} \mathcal{D}_{\alpha_1}(\sigma_1, \sigma^*)-\epsilon_1 \chi_1^{-} \mathcal{D}_{\beta_1}(\sigma_1, \sigma^*).
\end{aligned}
$$
where $\mathcal{D}_l(\sigma_1, \sigma^*) \triangleq (\sigma^*)^l+l (\sigma^*)^{l-1}(\sigma_1-\sigma^*)-\sigma_1^l$ with $\sigma_1 \triangleq \int_0^\infty x_1^T \Sigma_q x_1 dt$ and $\sigma^*\triangleq \int_0^\infty {x^*}^T\Sigma_qx^* dt$. To establish the required bounds, $e_1 \triangleq x_1-x^*$ is used as the trajectories resulting from deviations of the pursuer
\begin{equation}
\label{eq:deviation_dynamics}
\dot{e}_1=F e_1+B_1 (u-u^*)
\end{equation}
with $e_1(0)=0$ in either case. By the reverse triangle inequality in $\mathcal{L}_2$, one yields that
\begin{equation}
\label{eq:sigma_deviation_bound}
(\sqrt{\sigma_1}-\sqrt{\sigma^*})^2 \leq \int_0^\infty e_1^T\Sigma_q e_1\,dt.
\end{equation}

For any constant $0<l<1$, the concavity of the power function gives
$$
0\leq\mathcal D_l(\sigma_1,\sigma^*) \leq \max\{l,1-l\}(\sigma^*)^{l-1} (\sqrt{\sigma_1}-\sqrt{\sigma^*})^2.
$$
For $l=1$, $\mathcal D_1(\sigma_1,\sigma^*)=0$. Since $\sigma^*>h_c$, the definition of $\ell_1$ and (\ref{eq:sigma_deviation_bound}) yield
\begin{equation}
\label{eq:unified_remainder_bound}
\epsilon_1\chi_1^-\mathcal D_{\beta_1}(\sigma_1,\sigma^*) \leq \frac{\ell_1}{\lambda_{\max}(\Sigma_q)} \int_0^\infty e_1^T\Sigma_q e_1 dt.
\end{equation}

\textbf{Scenario 1.} Applying the $\mathcal{L}_2$-gain bound to \eqref{eq:deviation_dynamics} gives
\begin{equation}
\label{eq:scenario1_gain}
\begin{aligned}
\int_0^\infty e_1^T\Sigma_q e_1 dt \leq & \frac{ \lambda_{\max}(\Sigma_q) \lambda_{\max}(\Delta_S) \lambda_{\max}(S_1)}{f^2 \lambda_{\min}(\Delta_S)}\\
&\times \int_0^{\infty} (u-u^*)^T R (u-u^*) dt.
\end{aligned}
\end{equation}

Combining \eqref{eq:unified_remainder_bound} and \eqref{eq:scenario1_gain}, condition~(\ref{condition 13-c}) further implies
\begin{equation}
\label{eq:nash_remainder_energy}
\epsilon_1 \chi_1^{-} \mathcal{D}_{\beta_1}\left(\sigma_1, \sigma^*\right) \leq \int_0^{\infty} (u-u^*)^T R (u-u^*) dt.
\end{equation}

\textbf{Scenarios 2 and 3.} In Scenario~2, it follows that
\begin{equation}
\label{eq:scenario2_gain}
\begin{aligned}
\int_0^\infty e_1^T\Sigma_q e_1 dt \leq & \frac{\lambda_{\max}(\Sigma_q) D_2^2\lambda_{\max}(S_1)}{\lambda_{\min}^2(\mathcal Q_1-\mathcal Q_2)}\\
& \times \int_0^{\infty} (u-u^*)^T R (u-u^*) dt.
\end{aligned}
\end{equation}

By condition~(\ref{condition 23-b}), $\lambda_{\min}(\mathcal{Q}_1-\mathcal{Q}_2) >D_2\sqrt{\ell_S}$, combining (\ref{eq:scenario2_gain}) with (\ref{eq:unified_remainder_bound}) gives (\ref{eq:nash_remainder_energy}). 

For Scenario~3, the same argument, together with $P_1-P_2>0$ and condition~(\ref{condition 33}), yields the same bound in (\ref{eq:nash_remainder_energy}). Hence, (\ref{eq:nash_remainder_energy}) holds for the pursuer in all three scenarios. Since $\mathcal D_{\alpha_1}(\sigma_1,\sigma^*)\geq0$, the pursuer's index difference yields $J_1(u^*,v^*)\leq J_1(u,v^*)$.

For the evader, it follows that $\dot e_2 = Fe_2-B_2(v-v^*), \;\; e_2(0)=0$, where $e_2\triangleq x_2-x^*$, $\sigma_2\triangleq \int_0^\infty x_2^T\Sigma_qx_2\,dt$, and $v^*$ is evaluated along $x_2$. Repeating the preceding argument for the evader, one yields that
$$\epsilon_2\chi_2^-\mathcal D_{\beta_2}(\sigma_2,\sigma^*) \leq \int_0^\infty (v-v^*)^T\Pi(v-v^*) dt.
$$

Since $\mathcal D_{\alpha_2}(\sigma_2,\sigma^*)\geq0$, $J_2(u^*,v)-J_2(u^*,v^*) \leq 0$. It follows that the optimal controllers (\ref{CPT controller}) constitute a CPT-Nash equilibrium (\ref{CPT nash equilibrium}).

\section{Proof of Corollary~1} A stacking representation is applied for the equivalence of two quadratic forms. Define the stacked matrix
$$
\mathcal{P} \triangleq \begin{bmatrix} P_1\\ P_2 \end{bmatrix}\in\mathbb R^{2n\times n}.
$$

Following the block multiplication (\ref{blocking matrix}), the coupled CPT-based AREs (\ref{CPT Riccati equation}) are equivalently expressed as
\begin{equation}
\label{stacked}
\mathcal{P}^T \mathcal{S}_1 \mathcal{P} = \mathcal{Q}_1, \; \mathcal{P}^T \mathcal{S}_2 \mathcal{P} = -\mathcal{Q}_2.
\end{equation}

(a) \textbf{Scenario 1.} Since $S_1>0$, $S_2 >0$ and $\Delta_S >0$, by the Schur complement, one yields that $\mathcal{S}_1 > 0$. Moreover, the condition (\ref{condition 12-a}) implies that $\mathcal{Q}_1>0$. For any symmetric stabilizing solution, the column-orthonormal matrix $U_1\in\mathbb{R}^{2n\times n}$ is defined as $U_1 \triangleq \mathcal S_1^{1/2}\mathcal{P} \mathcal{Q}_1^{-1/2}$ with $U_1^TU_1=I_n$.

Substituting $\mathcal P$ and $U_1$ into (\ref{stacked}) implies (\ref{U}). The stacked solution can be written as $\mathcal{P} = \mathcal{S}_1^{-1/2} U_1 \mathcal{Q}_1^{1/2}$, whose blocks yield (\ref{Solution 1}). The converse follows directly from (\ref{stacked}). Hence, if these blocks are symmetric positive definite and the associated closed-loop matrix is Hurwitz, they constitute a symmetric stabilizing solution.

(b) \textbf{Scenario 2.} The proof follows the same argument as that for Scenario~1 and is therefore omitted to avoid repetition.

(c) \textbf{Scenario 3.} In this case, the coupled CPT-based AREs reduce to (\ref{ARE3}). For any $C=C^T>0$, choosing (\ref{Solution 3}) gives $P_1-P_2=S^{-1/2} \left(S^{1/2}\mathcal{Q}_1S^{1/2}\right)^{1/2}S^{-1/2}$, and the coupled CPT-based AREs are satisfied. The corresponding closed-loop matrix $F$ satisfies $S^{-1/2}FS^{1/2} =-\left(S^{1/2}\mathcal{Q}_1S^{1/2}\right)^{1/2}<0$. Conversely, any symmetric stabilizing solution satisfies $(P_1-P_2)S(P_1-P_2)=\mathcal{Q}_1$, since the associated matrix $F=-S(P_1-P_2)$ is Hurwitz, one necessarily has $P_1>P_2$. Taking $C=P_2>0$ uniquely gives the solution of the preceding equation by (\ref{Solution 3}).

Thus, all symmetric stabilizing solutions in Scenario~3 have the parameterized form. The proof is complete.

\bibliographystyle{IEEEtran}
\bibliography{ref}

\end{document}